\documentclass[prb, preprint,showpacs,preprintnumbers,amsmath,amssymb]{revtex4}

\usepackage{graphicx}% Include figure files
\usepackage{dcolumn}% Align table columns on decimal point
\usepackage{bm}% bold math

\begin{document}

\preprint{APS/123-QED}

\title{Acceleration of Phase Separation in Fe-base Ternary Alloys}% Force line breaks with \\

\author{Yoshihiro Suwa, Kanako Goto, Kazumi Ochi}
\author{Yoshiyuki Saito}%
 \email{ysaito@mn.waseda.ac.jp}
\affiliation{%
Department of Materials Science and Engineering, Waseda University\\
3-4-1 Okubo, Shinjuku-ku, Tokyo 169-8555, Japan
}%

\date{\today}% It is always \today, today,

\begin{abstract}
Mechanism for acceleration of phase separation in Fe-base ternary alloys was investigated with use of a model based on the Cahn-Hilliard equation. Behavior of the minor element in an Fe-base ternary alloy along the trajectory of the peak of the major element is dependent on the sign of the second derivative of the chemical free energy with respect to the concentrations of the major and minor elements. However, the concentration of the major element along the trajectory of its peak top increases with time regardless of the sign of the second derivative of the chemical free energy. The addition of a substitutional element to an Fe-base binary alloy with composition within the spinodal region was found to accelerate phase separation
\end{abstract}

\maketitle

\section{Introduction}

Control of microstructural evolutions associated with phase separation is important for materials design of Fe-based alloys.
Nanoscale microstructures resulting from heat treatment sometimes deteriorate properties of Fe-based alloys. For example, the ferrite phase in duplex stainless steels is thermomechanically unstable at service temperatures and hardens and embrittles due to the formation of modulated structure via phase separation \cite{hyde}. 

\begin{figure}
\includegraphics[width=17cm, height=10cm]{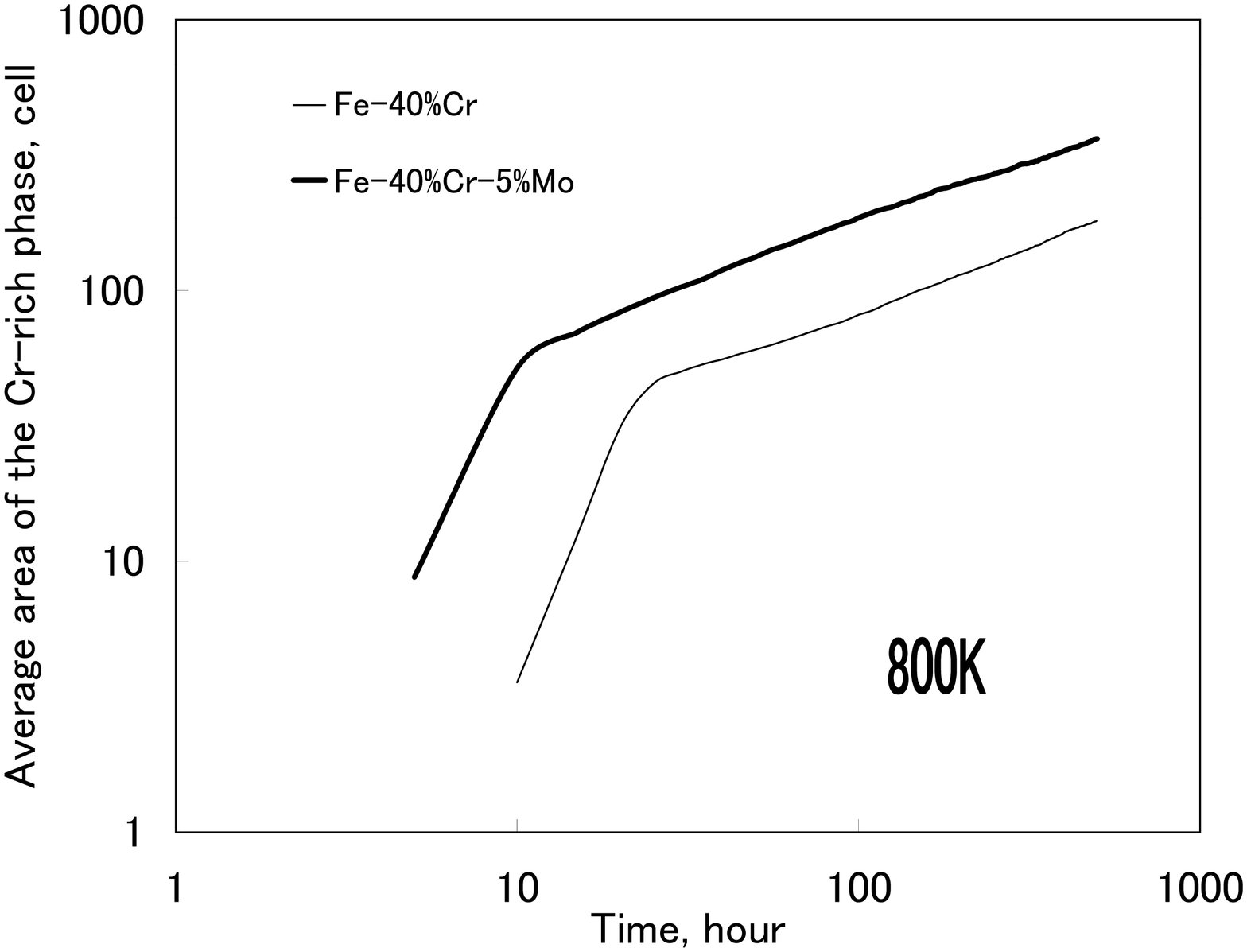}% Here is how to import EPS art
%\vspace{6cm}
\caption{\label{fig:eps1}Variation in the average areas of Cr-rich phase in Fe-Cr binary and Fe-Cr-Mo ternary alloys}
\end{figure}

\begin{figure}
\includegraphics[width=17cm, height=10cm]{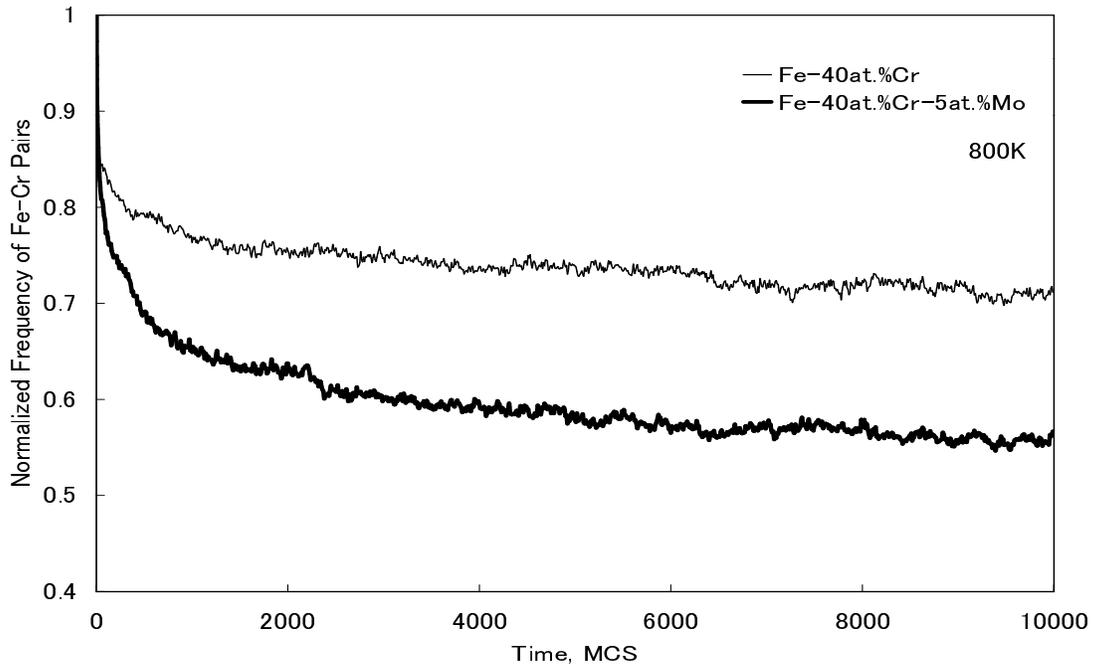}% Here is how to import EPS art
%\vspace{4cm}
\caption{\label{fig:eps2}Variation in the number of Fe-C pairs normalized by the intitial value with time in Fe-Cr binary and Fe-Cr-Mo ternary alloys}
\end{figure}

Behavior of phase separation in ternary alloys \cite{koyama} is known to be different from that of binary alloys. According to  
numerical simulations of phase separation in Fe-Cr binary  and Fe-Cr-Mo ternary alloys 
by Suwa and Saito \cite{suwa}, the addition of Mo to the Fe-Cr binary alloys accelerates decomposition of Cr. Figure \ref{fig:eps1} shows variations in the average areas of the Cr-rich phase in an Fe-40at.\%Cr binary alloy and an Fe-40at.\%Cr-5at.\%Mo ternary alloy with time. This figure indicates that the decomposition of Cr in the ternary alloy is faster than that in the binary alloy. The Monte Carlo simulation result \cite{goto1} of phase separation in Fe-Cr-Mo ternary alloys is shown in Fig.~\ref{fig:eps2}. The decreasing rate in the number of Fe-Cr pairs, which is closely related to the rate of phase separation, in the Fe-40at.\%Cr-5at.\%Mo ternary alloy is higher than that in the Fe-40at.\%Cr binary alloy.

This paper deals with the acceleration of phase decomposition process induced by the addition of a substitutional element such as Mo to an Fe-base binary alloy with a miscibility gap.  Mechanism for acceleration of phase separation is investigated by using a theory on asymptotic behavior of substitutional elements in ternary alloys proposed by the present authors \cite{ys3}.
\clearpage

\section{Fundamental Equations}
The Cahn-Hilliard equation for an Fe-X-Y ternary alloy is given by
\begin{eqnarray}
\frac{\partial c_X}{\partial t}&=&M_X\frac{\partial ^2}{\partial x^2}
\left(\frac{\partial f'}{\partial c_X} -K_X\frac{\partial ^2c_X}{\partial x^2}
-L_{XY}\frac{\partial ^2c_Y}{\partial x^2}\right) \label{eq:ps6} \\
\frac{\partial c_Y}{\partial t}&=&M_Y\frac{\partial ^2}{\partial x^2}
\left(\frac{\partial f'}{\partial c_Y}
-L_{YX}\frac{\partial ^2c_X}{\partial x^2} 
 -K_Y\frac{\partial ^2c_Y}{\partial x^2}\right)
\label{eq:ps7}
\end{eqnarray}
where $f'$ is local chemical free energy of the alloy with uniform composition, $K_X$, $K_Y$ and $L_{XY}$ are the gradient energy coefficients and $c_X(x,t)$ and $c_Y(x,t)$ are concentration fields of the X and the Y elements, respectively. Here we assumed the mobilities of the X and Y elements, $M_X$ and $M_Y$, are not dependent on their positions in the space. Equations (\ref{eq:ps6}) and (\ref{eq:ps7}) yield

\begin{widetext}
\begin{eqnarray}
\frac{\partial c_X}{\partial t}&=&M_X\left[\frac{\partial^2f'}{\partial c_X^2}\frac{\partial^2c_X}{\partial x^2}
+\frac{\partial^2f'}{\partial c_X\partial c_Y}\frac{\partial^2c_Y}{\partial x^2}
+2\frac{\partial^3f'}{\partial c_X^2\partial c_Y}\frac{\partial c_X}{\partial x}\frac{\partial c_Y}{\partial x} 
+\frac{\partial^3f'}{\partial^3c_X}\left(\frac{\partial c_X}{\partial x}\right)^2 \right.
\nonumber \\
&+&\left.\frac{\partial^3f'}{\partial c_X\partial^2c_Y}\left(\frac{\partial c_Y}{\partial x}\right)^2
-K_X \frac{\partial^4c_X}{\partial x^4}-L_{XY} \frac{\partial^4c_Y}{\partial x^4}\right] \\
\frac{\partial c_Y}{\partial t}&=&M_Y\left[
\frac{\partial^2f'}{\partial c_X\partial c_Y}\frac{\partial^2c_X}{\partial x^2}
+\frac{\partial^2f'}{\partial c_Y^2}\frac{\partial^2c_Y}{\partial x^2}
+2\frac{\partial^3f'}{\partial c_X\partial c_Y^2}\frac{\partial c_X}{\partial x}\frac{\partial c_Y}{\partial x} 
+\frac{\partial^3f'}{\partial c_X^2\partial c_Y}\left(\frac{\partial c_X}{\partial x}\right)^2 \right.
\nonumber \\
&+&\left.\frac{\partial^3f'}{\partial c_Y^3}\left(\frac{\partial c_Y}{\partial x}\right)^2
-L_{YX}\frac{\partial^4c_X}{\partial x^4}-K_Y \frac{\partial^4c_Y}{\partial x^4}
\right]
\label{eq:ps9}
\end{eqnarray}

Now we deal with asymptotic behavior of the X and Y elements at peaks of $c_X$.
Functions $G_X$ and $G_Y$ are difined as

\begin{eqnarray}
G_X\left(t,x,c_X, c_Y,\frac{\partial c_X}{\partial x},\frac{\partial c_Y}{\partial x},\frac{\partial c_X^2}
{\partial x^2},\frac{\partial c_Y^2}{\partial x^2},\frac{\partial^4c_X}{\partial x^4},\frac{\partial^4c_Y}
{\partial x^4}\right)
\equiv M_X\left[\frac{\partial^2f'}{\partial c_X^2}\frac{\partial^2c_X}{\partial x^2}
+\frac{\partial^2f'}{\partial c_X\partial c_Y}\frac{\partial^2c_Y}{\partial x^2}
\right. \nonumber \\
\left.
+2\frac{\partial^3f'}{\partial c_X^2\partial c_Y}\frac{\partial c_X}{\partial x}\frac{\partial c_Y}{\partial x}
+\frac{\partial^3f'}{\partial c_X^3}\left(\frac{\partial c_X}{\partial x}\right)^2+\frac{\partial^3f'}
{\partial c_X\partial^2c_Y}\left(\frac{\partial c_Y}{\partial x}\right)^2
-K_X\frac{\partial^4c_X}{\partial x^4}-L_{XY}\frac{\partial^4c_Y}{\partial x^4}\right]  \\
G_Y\left(t,x,c_X, c_Y,\frac{\partial c_X}{\partial x},\frac{\partial c_Y}{\partial x},\frac{\partial^2c_X}{\partial x^2},\frac{\partial^2c_Y}{\partial x^2},\frac{\partial^4c_X}{\partial x^4},\frac{\partial^4c_Y}{\partial x^4}\right)
\equiv M_Y\left[\frac{\partial^2f'}{\partial c_X\partial c_Y}\frac{\partial^2c_X}{\partial x^2}
+\frac{\partial^2f'}{\partial c_Y^2}\frac{\partial^2c_Y}{\partial x^2}
\right. \nonumber \\
\left.
+2\frac{\partial^3f'}{\partial c_X\partial c_Y^2}\frac{\partial c_X}{\partial x}\frac{\partial c_Y}
{\partial x}+\frac{\partial^3f'}{\partial c_X^2\partial c_Y}\left(\frac{\partial c_X}{\partial x}\right)^2
+\frac{\partial^3f'}{\partial c_Y^3}\left(\frac{\partial c_Y}{\partial x}\right)^2
-L_{YX}\frac{\partial^4c_X}{\partial x^4}-K_Y\frac{\partial^4c_Y}{\partial x^4}\right]
\end{eqnarray}
These function satisfy the following conditions:
\begin{eqnarray}
G_X\left(t,x_p,c_X, c_Y,0,\frac{\partial c_Y}{\partial x},0,\frac{\partial^2c_Y}{\partial x^2},
\frac{\partial^4c_X}{\partial x^4},\frac{\partial^4c_Y}{\partial x^4}\right)&=& \nonumber  \\
M_X\left[\frac{\partial^2f'}{\partial c_X\partial c_Y}\frac{\partial^2c_Y}{\partial x^2}
+\frac{\partial^3f'}{\partial c_X\partial c_Y^2}\left(\frac{\partial c_Y}{\partial x}\right)^2
\right.&-&\left.K_X\frac{\partial^4c_X}{\partial x^4}-L_{XY}\frac{\partial^4c_Y}{\partial x^4}\right]
 \quad \quad \\
G_Y\left(t,x_p,c_X, c_Y,0,\frac{\partial c_Y}{\partial x},0,\frac{\partial^2c_Y}{\partial x^2},\frac{\partial^4c_X}{\partial x^4},\frac{\partial^4c_Y}{\partial x^4}\right)&=& \nonumber  \\
M_Y\left[\frac{\partial^2f'}{\partial c_Y^2}\frac{\partial^2c_Y}{\partial x^2}
+\frac{\partial^3f'}{\partial c_Y^3}\left(\frac{\partial c_Y}{\partial x}\right)^2\right.&-&\left.L_{YX}\frac{\partial^4c_X}{\partial x^4}-K_Y\frac{\partial^4c_Y}{\partial x^4}\right] \quad \quad 
\end{eqnarray}
At a peak position $p(x_p,t)$
\begin{equation}
\frac{\partial c_X}{\partial x}=0 \quad \frac{\partial^2c_X}{\partial x^2}<0
\end{equation}
Applying the mean value theorem of differential calculus for compound fuction \cite{cour},
we obtain the following equation for an intermediate value $\zeta$ in the open interval $(\partial^2c_X/\partial x^2,0)$
\begin{eqnarray}
\frac{d c_X}{dt}(x_p,t)
&=&M_X\left[\frac{\partial^2f'}{\partial c_X^2}\frac{\partial^2c_X}{\partial x^2}(\zeta,t)
+\frac{\partial^2f'}{\partial c_X\partial c_Y}\frac{\partial^2c_Y}{\partial x^2}
+\frac{\partial^3f'}{\partial c_X\partial c_Y^2}\left(\frac{\partial c_Y}{\partial x}\right)^2 
-K_X\frac{\partial^4c_X}{\partial x^4}-L_{XY}\frac{\partial^4c_Y}{\partial x^4}\right]\nonumber \\
\label{eq:ps19} \\
\frac{d c_Y}{dt}(x_p,t)
&=&M_Y\left[\frac{\partial^2f'}{\partial c_X\partial c_Y}\frac{\partial^2c_X}{\partial x^2}(\zeta,t)
+\frac{\partial^2f'}{\partial c_Y^2}\frac{\partial^2c_Y}{\partial x^2}
+\frac{\partial^3f'}{\partial c_Y^3}\left(\frac{\partial c_Y}{\partial x}\right)^2 -L_{YX}\frac{\partial^4c_X}{\partial x^4}-K_Y\frac{\partial^4c_Y}{\partial x^4}\right]
\label{eq:ps20}
\end{eqnarray}
\end{widetext}

\section{Mechanism for Acceleration of Phase Separation in Ternary Alloys}
\subsection{Asymptotic Behavior of Substitutional Elements in Fe-base Ternary alloys}
Here we consider the case in which the concentration of the X element
in an Fe-X-Y ternary alloy, $c_X$
and the concentration of the Y element, $c_Y$ satisfy the following conditions
\begin{equation} 
c_X>c_Y, \quad
\frac{\partial^2f'}{\partial c_X^2}<0, \quad  \frac{\partial^2f'}{\partial c_Y^2}>0
\end{equation}

The mechanism for bifurcation  formation of peaks can be explained by a theory proposed by the present author\cite{ys3}.
Equation (\ref{eq:ps20}) along a peak top of $c_X$ can be approximated by
\begin{eqnarray}
\frac{d c_Y}{d t}(x_p,t)\simeq M_Y\frac{\partial^2f'}
{\partial c_X \partial c_Y}\frac{\partial^2c_X}{\partial x^2}
\label{eq:ps23}
\end{eqnarray}
From Eq.(\ref{eq:ps23}) it is indicated that the behavior of the Y elements at
a peak position of the X depends on the sign of $f'_{XY}\equiv \partial^2f'/\partial c_X \partial c_Y$.
With use of the regular solution model, we have
\begin{eqnarray}
f'_{XY}&\equiv &\frac{\partial^2f'}{\partial c_X\partial c_Y} \nonumber \\
&=&\Omega_{XY}-\Omega_{FeX}-\Omega_{FeY}+RT\frac{1}{1-c_X-c_Y}
\label{eq:ps13}
\end{eqnarray}
where $R$ is the gas constant, $T$ is the absolute temperature, $\Omega_{FeX}$, $\Omega_{FeY}$ and $\Omega_{XY}$
are interaction parameters.
If $f'_{XY}<0$, then we have
\begin{equation}
\frac{dc_Y(x_p,t)}{dt}>0
\label{eq:ps25}
\end{equation}
Peaks of $c_Y$ will be formed at the same position of the peak tops
of $c_X$. This is 
a phase separation induced modulated structure.
If $dc_X/dt>0$ and $dc_Y/dt>0$ then
\begin{equation}
\frac{df'_{XY}}{dt}>0
\label{eq:ps27}
\end{equation}
The sign of $f'_{XY}$ may
change from negative to positive in lower temperatures
at which the equilibrium concentration of the X element is high.
This indicates that bifurcation of peaks will occur at the 
later stage of phase decomposition.

Now let us consider asymptotic behavior of the X element at peak tops of $c_X$. Once
peaks or bottoms of $c_Y$ form at peak tops of $c_X$, the third term of Eq.(\ref{eq:ps19})
is zero. The fourth order derivative terms attribute to the interfacial energies of the X and Y elements in the Fe matrix. The contribution of these terms at the peaks of $c_X$ is assumed  to be smaller than the first and the second terms.
Then at peaks of $c_X$, Eq.(\ref{eq:ps19}) can be approximated by
\begin{eqnarray}
\frac{d c_X}{d t}(x_p,t)\simeq M_X\left[\frac{\partial^2f'}{\partial c_X^2}\frac{\partial^2c_X}{\partial x^2}
+\frac{\partial^2f'}{\partial c_X\partial c_Y}\frac{\partial^2c_Y}{\partial x^2}\right]
\label{eq:ps29}
\end{eqnarray}
The first term in the right hand side of the above equation describes the phenomena of uphill diffusion, which is the major feature of phase decomposition in binary alloys. The second term represents the effect of the element Y on 
variation in $c_X$ induced by the addition of the Y elements. 

\subsection{Acceleration of Phase Decomposition}
Because of a strong repulsion between Fe and X atoms phase decomposition in an Fe-X binary alloy is initiated by a continous process at spinodal region. Variation of the peak top behavior of the X element induced by the addition of   Y element will be analyzed with use of Eq.(\ref{eq:ps29}). 
The interaction parameter of the X and Y elements in an Fe-X-Y ternary alloy is difined as
\begin{equation}
W_{XY}=\Omega_{XY}-\Omega_{FeX}-\Omega_{FeY}
\label{ps30}
\end{equation}

\paragraph{$f'_{XY}<0$(case 1)} 
If the value of $W_{XY}$ is negative and its absolute value  is larger than $RT/c_{Fe}$, $f'_{XY}$ is negaive at the initial time. The amplitude of $c_Y$ along the peaks of $c_X$ increases with time. Then peaks of $c_Y$ form at peaks of $c_X$. It follows that the sign of the second term in Eq.(\ref{eq:ps29}) is positive. This accelerates decomposition of the X element. At a low temperature, the value of $RT/c_{Fe}$ becomes larger than $|W_{XY}|$ at the later stage. Then the value of $f'_{XY}$ changes from negative to positive and the bifurcation of peaks of $c_Y$ will occur. Because the signs of both $\partial^2c_Y/\partial x^2$ and $f'_{XY}$ are positive, the positive sign of the second term will not change at the later stage. If the value of $f'_{XY}$ is negative at the initial time, phase separation in Fe-X-Y ternary alloys is accelerated by the addition of the Y element.

\paragraph{$f'_{XY}>0$(case 2)}
If the value of $W_{XY}$ is positive, then the value of $f'_{XY}$ is positive.
From Eq.(\ref{eq:ps23}) it follows that the amplitude of $c_Y$ along the peak top of $c_X$ decreases with time. Bottoms of $c_Y$ will form at peaks of $c_X$. The amplitude of $c_X$ along its peaks increases with time because the signs of both $\partial^2c_Y/\partial x^2$ and $f'_{XY}$ are positive in this case. The phase decomposition of the X element is also accelerated in the case that the value of $f'_{XY}$ is positive.
\\

\begin{figure}
\includegraphics[width=17cm, height=22cm]{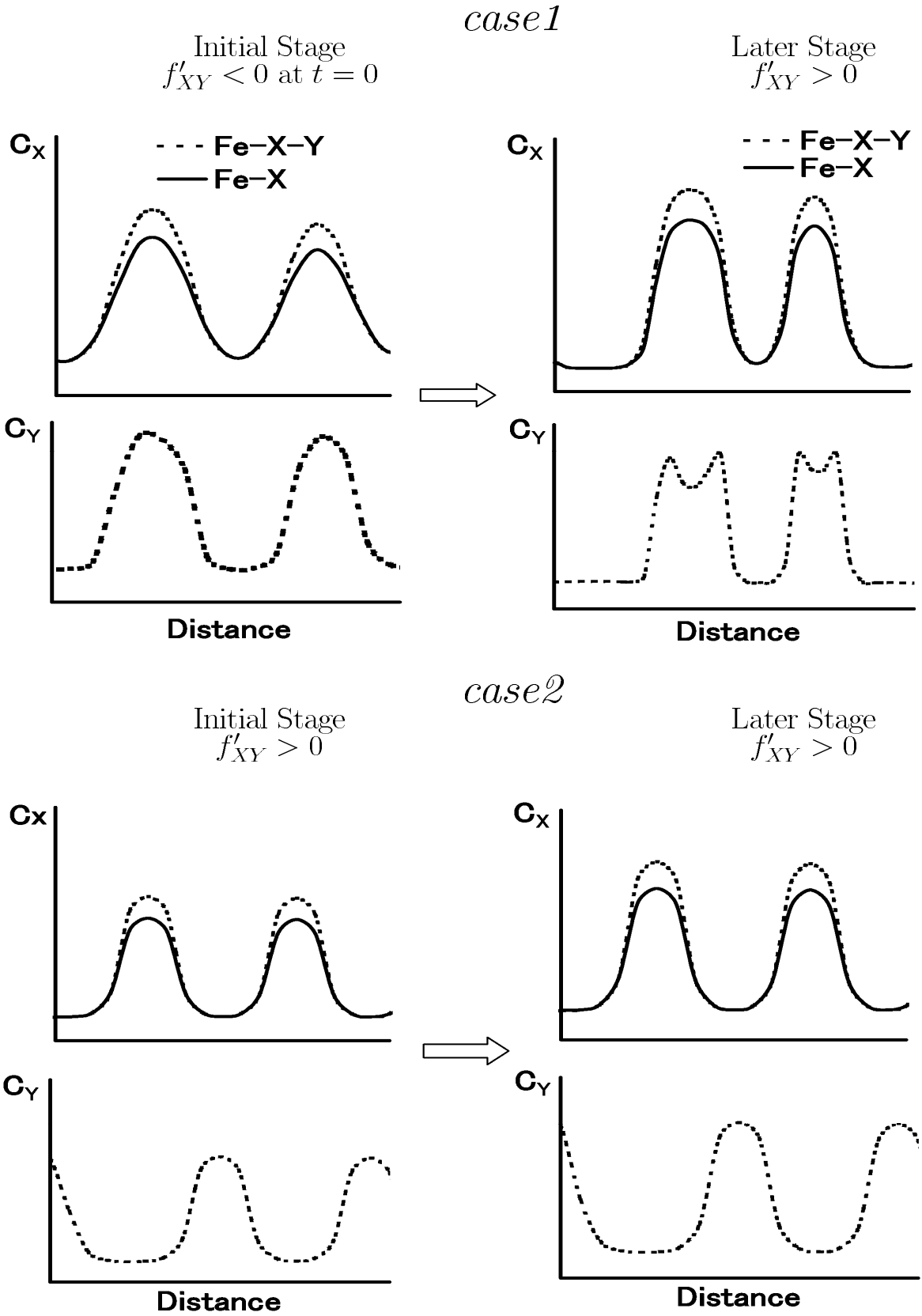}% Here is how to import EPS art
%\vspace{10cm}
\caption{\label{fig:eps3}Schematic illustration of the mechanism for acceleration of phase separation in an Fe-X binary alloy induced by the addition of Y element}
\end{figure}

Acceleration of decomposition of the X element induced by the addition of the Y element to binary alloys with miscibility gaps 
occurs regardless of the sign of $f'_{XY}$.
Figure \ref{fig:eps3} shows schematic illustration of the mechanism for acceleration of phase separation in an Fe-X binary alloy induced by the addition of Y element.

\section{Conclusion}
We proposed a mechanism for acceleration of phase separation induced by the addition of the third element to Fe-base binary alloys with miscibility gaps. The addition of Y element to an Fe-X binary alloy with composition within the spinodal region accelerates the decomposition of the X element regardless of the sign of the second derivative of the chemical free energy, $f'$, with respect to the concentrations of X and Y, $c_X$ and $c_Y$. While the sign of the second
 derivative, 
$f'_{XY}\equiv \partial^2f'/\partial c_X \partial c_Y$, 
controls behavior of the Y element along the trajectory of the peak top of $c_X$. This mechanisms for acceleration of phase separation in Fe-X-Y ternary alloys is active in both cases that there is a repulsive interaction between the X and Y elements and that the interaction between the X and Y elements is attractive.

An effective way to prevent acceleration of phase transformation will be to minimize the absolute value of $f'_{XY}$.

\begin{acknowledgments}
The authors would like to acknowledge the continuing guidance and encouragement of ProfessorAkihiko Kitada. 
\end{acknowledgments}


\begin{thebibliography}{99}
\bibitem{hyde}
J.M. Hyde, A.Cerezo, M.K. Killer and G.U.W. Smith, Appl. Surf. Sci., {\bf 76/77}, 233(1994)
\bibitem{koyama}
T. Koyama, T. Kozakai and T.Miyazaki: in {\it Proceeding of the International Conference on Solid-Solid Phase Transformation '99}, (The Japan Institute of Metals, Sendai, Japan, 1999), edited by M. Koiwa, K. Otsuka and T. Miyazaki, p.733
\bibitem{suwa}
Y. Suwa and Y. Saito: CAMP-ISIJ, {\bf 14}, 1206(2001)
\bibitem{goto1}
K. Goto: Diploma Thesis, Dept. of Mater. Sci. \& Eng., Waseda University, (2001)
\bibitem{ch1}
J.W. Cahn and J.E. Hilliard: J. Chem. Phys. {\bf 28}, 258(1958)
\bibitem{ch2}
J.W. Cahn :Acta Metall., {\bf 9}, 795(1961)
\bibitem{ys3}
Y. Saito, Y. Suwa, K. Ochi, T. Aoki, K.Goto and K. Abe: submitted to J. Phys. Soc. Japan, (2001)
\bibitem{cour}
R. Courant and F. John: {\it Introduction to Calculas and Analysis I}, (Springer-Verlag, New York, 1989)
\end{thebibliography}
\end{document}